\documentclass[twocolumn,showpacs,preprintnumbers,amsmath,amssymb]{revtex4}

\usepackage{graphicx}

\newcommand{\Eref}[1]{eq.(\ref{#1})}
\newcommand{\Fref}[1]{Fig.\ref{#1}}
 \usepackage{bm}
 \usepackage{afterpage}

\begin{document}

\preprint{}

\title{Scanning tunneling spectroscopic evidence of crossover transition \\ in the two-impurity
Kondo problem}
\author{Emi Minamitani}
 \altaffiliation[]{Department of Precision Science \& Technology and Applied Physics Graduate School of  Engineering Osaka University.}
\author{Hiroshi Nakanishi}
\author{Hideaki Kasai}
 \email{kasai@dyn.ap.eng.osaka-u.ac.jp}
\affiliation{
Department of Precision Science \& Technology and Applied Physics, Osaka University Suita, Osaka 565-0871 JAPAN }%

\author{Wilson Agerico Di\~{n}o}
\affiliation{ Department of Physics, Osaka University, Toyonaka,Osaka 560-0043,
Japan }%

\date{\today}

\begin{abstract}
We calculate
the differential conductance ({\it dI/dV}) corresponding to scanning tunneling
spectroscopy (STS) measurements for two magnetic atoms adsorbed on a metal
surface with the aid of the numerical renormalization group
(NRG) technique. We find that the peak structure of the {\it dI/dV} spectra near
the Fermi level changes gradually as a function of the adatom separation and the
coupling between the adatoms and the metal surface conduction band.
When the coupling becomes small, the peak disappears and, instead, a dip
structure appears near the Fermi level. This dip structure is the manifestation of the strong
antiferromagnetic correlation between the localized spins. The gradual change
of the {\it dI/dV} structure from a peak structure to a dip structure originates
from the crossover transition in the two impurity Kondo problem.
\end{abstract}

\maketitle

\section{Introduction}
\begin{figure}[htp]
\begin{center}
  \includegraphics[keepaspectratio=true,width=80mm]{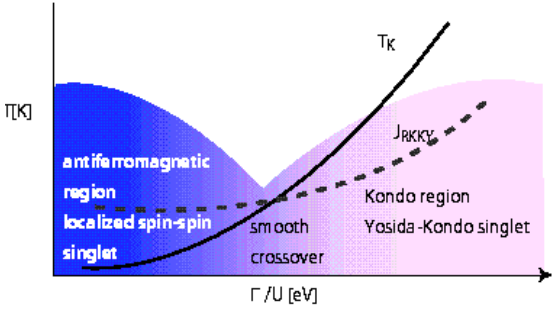}
  \caption[labelInTOC]{Diagram of the smooth crossover in the two impurity Kondo effect.
  $T$ is the temperature. $T_K$ is the Kondo temperature and $J_{RKKY}$ is the
  RKKY coupling constant.}
  \label{fig:Tk_TRKKY.eps}
\end{center}
\end{figure}

~~The two-impurity Kondo problem has long been extensively investigated.
Generally, the low temperature physics depends on the ratio between the
RKKY coupling constant $J_{\mathrm {RKKY}}$ and the one-impurity Kondo temperature $T_{\mathrm K}$ \cite{Jayaprakash1981}.
In the limit of strong ferromagnetic RKKY coupling ($-J_{\mathrm {RKKY}}>>
T_{\mathrm K}$), a two stage Kondo effect occurs: First the localized spins are partially compensated as temperature
$T$ decreases, and then completely suppressed as $T$ goes to zero. In the
limit of strong antiferromagnetic RKKY coupling ($J_{\mathrm {RKKY}}>>
T_{\mathrm K}$), the localized spins are locked into a spin-singlet state (the
antiferromagnetic region), and the Kondo effect plays a minor role. In
the $T_{\mathrm K} >> |J_{\mathrm {RKKY}}|$ region (the Kondo region), the RKKY
interaction plays a minor role. \\
~~The scenarios in the corresponding limits are very
reasonable, but the physics in the region at intermediate values of $J_{\mathrm
{RKKY}}$ is nontrivial. Studies carrying out numerical renormalization
group (NRG) diagonalization on the electron-hole (e-h) symmetric Hamiltonian found
a critical point which separates the Kondo region and the antiferromagnetic
region \cite{Jones1988,Jones1989}. However, it is now apparent that the critical point
results from the e-h symmetry of the model
Hamiltonian\cite{Affleck1995,Sakai1992a}. In general, e-h symmetry is broken
(e.g., energy dependence in tunneling matrix elements) and the quantum phase
transition is replaced by a crossover transition \cite{Silva1996}. \\ ~~The
magnetic atom dimer on a nonmagnetic metal surface is a classical example of a two-impurity Kondo system. In this system, the strength and sign
of the RKKY interaction can be adjusted by changing the adatom separation.
Using STS, the Kondo effect can be observed through the sharp peak structure
near Fermi level, which corresponds to the Yosida-Kondo singlet
\cite{Kasai2000,Kawasaka1999,Shimada2003,Dino2006,Madhavan1998,Heinrich2004,Crommie1993,Minamitani,Minamitani2009}.
Thus, we would expect to observe the interference between the Kondo effect and
the spin ordering effect of the RKKY interaction through the STS spectra.
Experimentally, the STS spectra of a Co dimer on a Cu(100) surface vary with the change of the adatom separation \cite{Wahl2007},
which is expected to result from the two impurity Kondo problem.
However, theoretical studies specific to STS measurements are rare\cite{Minamitani, Minamitani2009, Merino2009}.
In our earlier studies, we confirmed that the ferromagnetic (antiferromagnetic)
interaction tends to sharpen (broaden) the {\it dI/dV} peak structure
\cite{Minamitani, Minamitani2009}. On the other hand, we find from {\it dI/dV}
calculations based on e-h symmetric Hamiltonian that such a broad peak does not
exist. These results indicate that the observed broad STS spectra are related to the crossover between the Kondo region
and the antiferromagnetic region.\\
~~Both $T_{K}$ and $J_{RKKY}$ strongly depend on the
ratio between the Coulomb interaction $U$ and the coupling between the adatom
and the metal surface conduction band, $\Gamma$. As schematically shown in
\Fref{fig:Tk_TRKKY.eps}, $T_K$ decays much faster than $J_{RKKY}$ as $\Gamma$ decreases.
 With small
$\Gamma$, we would expect that the antiferromagnetic RKKY interaction becomes
dominant.
In this situation,
adjusting the RKKY interaction by changing the adatom separation,
we can trace the crossover between the Kondo region and the antiferromagnetic region.
For this reason, we calculate the {\it dI/dV} spectra for several values
of $\Gamma$ and the adatom separation to show how the crossover transition in
the two impurity Kondo problem can be observed through the STS spectra.
\section{Model and Method}
\begin{figure}[htbp]
  \begin{center}
    \includegraphics[keepaspectratio=true,height=50mm]{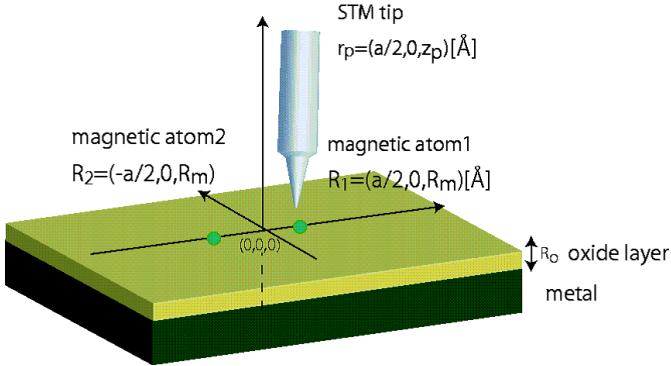}
  \end{center}
  \caption{STS observation of a magnetic dimer on a nonmagnetic metal surface. Distances between  the
  adatoms ($a$) , the tip apex and the metal surface ($z_p$) are given in {\AA}ngstr\"{o}ms [\AA].
  STM tip is placed directly above atom 1.
  }
  \label{fig:schematic1.eps}
\end{figure}

As experiments show\cite{Heinrich2004,Otte2008}, by covering the metal surface
with a decoupling layer, $\Gamma$ was suppressed, which leads to
a decrease in $T_K$. In the present study, we consider the model system shown in
\Fref{fig:schematic1.eps}.

Magnetic atom 2 is located at a distance of $a$ \AA\  with respect to magnetic atom 1. We assume
that the center of atom 1 is located at $\mathbf{R_1}=(a/2,0,R_m)$ \AA\  and
that of atom 2 is located at $\mathbf{R_2}=(-a/2,0,R_m)$ \AA. $R_m$ is the
radius of adatom. We set $R_m=0.4$ \AA\ \cite{Shimada2003}. The tip apex is at a distance of $z_{p}$ from the metal surface, directly above atom 1. $R_o$ is the thickness of the
decoupling layer and we set $R_o=5$\AA. The corresponding model Hamiltonian is given
by

\begin{eqnarray}
H&=&H_{A2}+H_{tip}+H_{mix}\nonumber \\
&=&\sum_{i\sigma}E_d d_{i\sigma}^\dagger d_{i\sigma}+\sum_{k,\sigma}E_k c_{k\sigma}^\dagger c_{k\sigma}+\sum_{ik\sigma}(V_{kdi}c_{k\sigma}^\dagger d_{i\sigma}+ h.c. )
\nonumber \\
&&+\sum_{i}U n_{i\uparrow} n_{i\downarrow}+\sum_{p,\sigma} E_p c^{\dagger}_{p\sigma} c_{p\sigma} \nonumber\\
&&+\sum_{li\sigma}(T_{pdi}c_{p\sigma}^\dagger d_{i \sigma} + h.c.)+\sum_{pk\sigma}(W_{pk}c_{p\sigma}^\dagger c_{k\sigma}+h.c.).
\label{H_gen}
\end{eqnarray}

\noindent Here $d_{i\sigma}^\dagger$, $c_{k\sigma}^\dagger$ and $c^{\dagger}_{p\sigma}$ correspond to
creation operators  for adatom $d$ electrons, metal surface conduction electrons, and tip electrons
with spin $\sigma$, respectively. $n_{i\sigma}=d^\dagger_{i\sigma}d_{i\sigma}$. $E_d$, $E_k$ and
$E_p$ correspond to the kinetic energies of adatom $d$ electrons, conduction electrons and tip
electrons, respectively. Adatom index $i=1,2$. $k$ corresponds to the metal surface electron
wavenumber, and $p$ corresponds to eigenstate quantum number of the tip electrons. $T_{pdi}$,
$W_{pk}$, and $V_{kdi}$ correspond to the tip-adatom, tip-surface, and adatom-surface electron
tunneling matrix elements, respectively. $U$ gives the on-site Coulomb repulsion on adatoms. We
approximate the coefficients in the Hamiltonian (\ref{H_gen}) as

\begin{eqnarray}
V_{kd1}&=&V_0 \exp(i\mathbf{k}\cdot a/2), V_{kd2}=V_0 \exp(-i\mathbf{k}\cdot a/2) , \label{V_kd} \\
W_{kp}&=&W_0 \exp(-(z_p-R_s+R_o)/\lambda)\exp(-i(\mathbf{k\cdot
r_p})),\nonumber \\
\\
 T_{pdi}&=&T_0\frac{\psi_{di}(\mathbf{r_p}-\mathbf{R_i})}{\psi_{di}(\mathbf{R_i}+\mathbf{R})}=T_0\Phi _{di},
\end{eqnarray}
and
\begin{eqnarray}
\Phi _{di}=\frac{\psi_{di}(\mathbf{r_p}-\mathbf{R_i})}{\psi_{di}(\mathbf{R_i}+\mathbf{R})}.
\label{coef}
\end{eqnarray}

\noindent $\psi_{di}$ corresponds to the $d$ electron orbital of adatom $i$. We approximate the tip
apex as a nonmagnetic metal sphere whose center is positioned at $\mathbf{r_p}=(a/2,0,z_p)$ with
radius $Rs$. $\mathbf{R}=(0,0,Rs)$ and $\mathbf{a}=(a,0,0)$. $W_0$ and $T_0$ correspond to the values
of the tunneling matrix elements when the tip is in contact with the surface or adatoms,
respectively. $V_0$ is the tunneling matrix element between a localized {\it d} electron and a metal
surface in the single impurity case.

Using non-equilibrium Green's function method\cite{Shimada2003,Rammer1986}, the electron current from
the STM tip to the surface can be written as

\begin{eqnarray}
I&=&\frac{2e}{h}\sum_{\sigma}\int d\omega (f_k-f_p) \Bigl\{2\pi T_0^2(\Phi_{d1}^2+\Phi_{d2}^2)\rho_p \Im G_{11\sigma}^r \nonumber \\
&+&4\pi T_0^2\Phi_{d1}\Phi_{d2}\rho_p \Im G_{12\sigma}^r -2\pi^2\rho_k\rho_p W_0^2 e^{-2(z_p-R_s)/\lambda} \nonumber\\
&-&4\pi T_0(\Phi_{d1}J_{vw}(r_1)+\Phi_{d2}J_{vw}(r_2))\rho_{p}\Re G_{11\sigma}^r \nonumber \\
&-&4\pi T_0(\Phi_{d1}J_{vw}(r_2)+\Phi_{d2}J_{vw}(r_1))\rho_{p}\Re G_{12\sigma}^r \nonumber \\
&-&2\pi(J_{vw}^2(r_1)+J_{vw}^2(r_2))\rho_{p}\Im G_{11\sigma}^r\nonumber \\
&-& 4\pi J_{vw}(r_1)J_{vw}(r_2)\rho_{p}\Im G_{12\sigma}^r \Bigr\}.
\label{J}
\end{eqnarray}

\noindent From Eq.~(\ref{J}) we can then obtain the corresponding differential conductance $dI/dV$.
In Eq.~(\ref{J}), the retarded Green's function $G_{ij\sigma}^r$ of the surface system corresponds to that of the two-impurity Anderson model.
$f_k$ and $f_p$ give the Fermi distribution functions for the surface and tip,
respectively.  $\rho_k$ and $\rho_p$ give the density of states of the
conduction electrons and the STM tip electrons, respectively. We define $\Phi_{di}$ and
$J_{vw}(r)$ as follows -
\begin{eqnarray}
J_{vw}(r)&=&\pi\rho_k J_0(k_Fr)V_0 W_0\exp(-(z_p-R_s+R_o)/\lambda).\nonumber \\
\end{eqnarray}

\noindent $\lambda$ is the decay constant of the surface electron wave function.
$J_0$ is the 0th order Bessel function.
Using these parameters, $\Gamma$ is defined as $\Gamma=\pi\rho_k V_0^2$.
At $T=0$~K, $dI/dV$ can be rewritten in the following simplified form -
\begin{eqnarray}
dI/dV(eV) & \propto & A_{11}\Im G_{11\sigma}^r(eV)+A_{12}\Im G_{12\sigma}^r(eV) \nonumber \\
&+&B_{11}\Re G_{11\sigma}^r(eV)+B_{12}\Re G_{12\sigma}^r(eV)
\label {dIdVform}
\end{eqnarray}
Coefficients such as $A_{11},B_{11}$ can be derived from the coefficients of
the Green's function in Eq.~(\ref{J}), i.e.,
\begin{eqnarray}
A_{11}&=&-2\pi T_0^2(\Phi_{d1}^2+\Phi_{d2}^2)\rho_p , \nonumber \\
&+&2\pi(J_{vw}^2(r_1)+J_{vw}^2(r_2))\rho_{p} ,  \label{A11} \\
A_{12}&=&-4\pi T_0^2\Phi_{d1}\Phi_{d2}\rho_p+ 4\pi
J_{vw}(r_1)J_{vw}(r_2)\rho_{p} , \label{A12}\\
B_{11}&=&4\pi
T_0(\Phi_{d1}J_{vw}(r_1)+\Phi_{d2}J_{vw}(r_2))\rho_{p}  \label{B11},
\\ B_{12}&=&4\pi T_0(\Phi_{d1}J_{vw}(r_2)+\Phi_{d2}J_{vw}(r_1))\rho_{p} .
\label {B_12}
\end{eqnarray}
The third and fourth terms are related to the Fano effect and makes the $dI/dV$ spectra asymmetric.
We show several values of each coefficient \Eref{dIdVform} in the case of
$\Gamma=0.022$eV at several adatom separation
in Table \ref{ABcoef} .

\begin{table}[htbp]
 \caption{Values of coefficients in Eq.(\ref{dIdVform}) in the case of
 $\Gamma=0.022$ eV at several adatom separations ($a$). }%
 \begin{center}
  \begin{tabular}{|c|c|c|c|c|}
    \hline
      $a$ (\AA) &  $A_{11}$  (eV) & $A_{12}$ (eV) & $B_{11}$  (eV)& $B_{12}$  (eV)  \\
    \hline
       5.0 &   -6.6225$ \times 10^{-4}$ &    -1.6792$ \times 10^{-9}$ &    5.3360$ \times 10^{-7}$ &
       1.7792$ \times 10^{-7}$ \\
    \hline
       6.0&   -6.6225$ \times 10^{-4}$ &   6.8125$ \times 10^{-10}$  & 5.3360$ \times 10^{-7}$
       &6.6775$ \times 10^{-8}$ \\
    \hline
       7.0 &  -6.6225$ \times 10^{-4}$  &  2.9035$ \times 10^{-11}$ &5.3360$ \times 10^{-7}$ &
       -3.5216$ \times 10^{-8}$ \\
    \hline
       8.0 & -6.6225$ \times 10^{-4}$& -4.6240$ \times 10^{-11}$&  5.3360$ \times 10^{-7}$
       &-1.1929$ \times 10^{-7}$ \\
    \hline
      9.0 &  -6.6225$ \times 10^{-4}$  &-7.1997$ \times 10^{-11}$&  5.3360$ \times 10^{-7}$&
      -1.7886$ \times 10^{-7}$\\
    \hline
  \end{tabular}
 \end{center}
 \label{ABcoef}
\end{table}
As can be seen from Table \ref{ABcoef}, in \Eref{dIdVform}, $A_{11}\Im
G_{11\sigma}^r(eV)$ is dominant and other parts play only minor roles.
This means that the decoupling layer suppresses the Fano
effect. To derive the relevant Green's function, we adopt the NRG technique
\cite{Krishna1980,Krishna1980a,Sakai1992a,Campo2005,Bulla2008}. In the present
study, we transform $H_{A2}$ in \Eref{H_gen} into two semi-infinite chains form
so as to be suitable for NRG calculation \cite{Campo2005,Silva1996,Minamitani,Minamitani2009}:

\begin{eqnarray}
H_{A2}&=&\sum_{n q=\pm}\epsilon_{nq} f^\dagger_{nq} f_{nq}+\sum_{n q=\pm}(t_{nq} f^\dagger_{nq} f_{n+1 q} +H.c.) \nonumber \\
&+& \sum_{q=\pm} E_d n_{dq}+U\sum_{i=1,2} n_{di\uparrow}n_{di\downarrow}
\nonumber \\ &+& \sqrt{\frac{2D\Gamma}{\pi}}\sum_{q=\pm}(\sqrt{\bar{\gamma}}f^\dagger_{0q} d_q +h.c. ).
\label{NRGform}
\end{eqnarray}

\noindent Here, $f_{nq}$ corresponds to the $n$th site of the conduction
electron part of the chain (Wilson chain) with the parity $q$. $q=+,-$ denotes
the even and odd parity states, respectively. $d_+=\frac{d_1+d_2}{\sqrt{2}}$ and
$d_-=\frac{d_1-d_2}{\sqrt{2}}$. $D$ gives the conduction electron band width. $\epsilon_{nq}$ and $t_{nq}$ are the corresponding matrix elements for the Wilson chain. When the metal surface conduction band is that for a two dimensional free electron,
\begin{eqnarray}
f_{0\pm}&=&\int^{D}_{-D} \sqrt{\gamma_\pm(\epsilon)}c_{\epsilon\pm} /\sqrt{\bar{\gamma_\pm}}d\epsilon, \\
\gamma_\pm(\epsilon)&=&\frac{1\pm J_0(ka)}{2} \label{gamma_pm},\\
\bar{\gamma_\pm}&=&\int^{D}_{-D} \gamma_\pm(\epsilon) d\epsilon.
\end{eqnarray}

\section{Numerical results and Discussions}

\begin{figure}[htb]
\begin{center}
  \includegraphics[keepaspectratio=true, height=130mm]{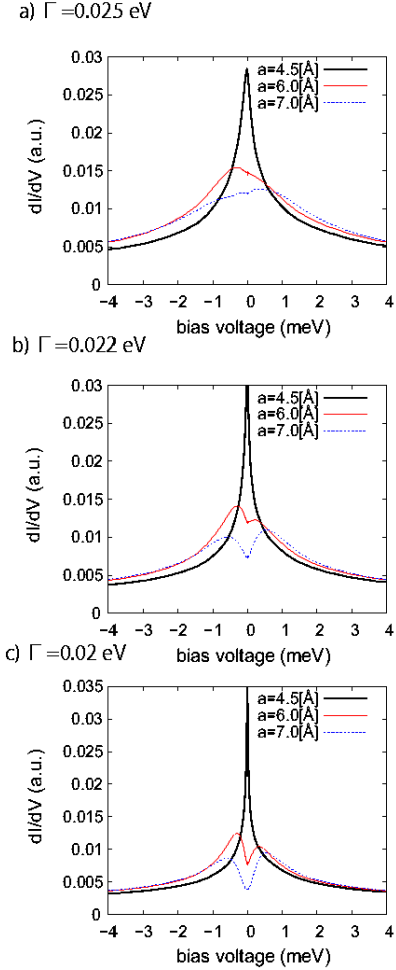}
  \caption[labelInTOC]{{\it dI/dV} calculation results at several adatom
  separations with a) $\Gamma=0.025$eV ($T_K$=6.64K), b)$\Gamma=0.022$eV ($T_K$=3.07K), and
  c) $\Gamma=0.02$eV ($T_K$=1.63K). The STM tip is placed at $z_p=4.0$\AA. We
  set $R_S=3.0$\AA~ and $W_0=T_0=0.02$eV.}
  \label{fig:contour_line.eps}
\end{center}
\end{figure}

The two dominant parameters $T_{K}$ and $J_{RKKY}$ strongly depend on the value
of $\Gamma$. $T_K$ decays much faster than $J_{RKKY}$ as $\Gamma$ decreases.
The values of $\Gamma$ (therefore
$T_K$) depend on the thickness, the surface condition and the type of the
decoupling layer. In the present study, we set the thickness of the decoupling
layer to 5\AA\ \cite{Heinrich2004}. In such system, the value of $T_K$ ranges
from 3 to 6K. Thus, we calculate the {\it dI/dV} spectra with
$\Gamma=0.025,0.022$, and $0.02$eV at several adatom separations. (From \cite{Yoshimori1986},
corresponding $T_K$ is estimated at 6.64, 3.07,and 1.63K, respectively) In
\Fref{fig:contour_line.eps}, we show the calculation results for the {\it dI/dV}
spectra.
At $a=5.0$\AA~ and $a=9.0$\AA, there is a sharp peak structure near the Fermi
level. With these adatom separations, the RKKY interaction is weak antiferromagnetic and the
Kondo effect is dominant\cite{Minamitani2009}. The sharp peak corresponds to the Yosida-Kondo
resonance. With our settings, the antiferromagnetic RKKY interaction becomes
largest around $a=7.0$\AA. As the adatom separation becomes close to 7.0\AA, the
peak structure changes gradually.
When $\Gamma=0.025$eV, the {\it dI/dV} spectra broaden as $a$ becomes close to
7.0\AA. However, when $\Gamma = 0.022$eV, a dip structure appears near the Fermi level.
This dip structure develops to a deeper one as $\Gamma$ decreases.

\begin{figure}[htp]
\begin{center}
  \includegraphics[keepaspectratio=true, height=100mm]{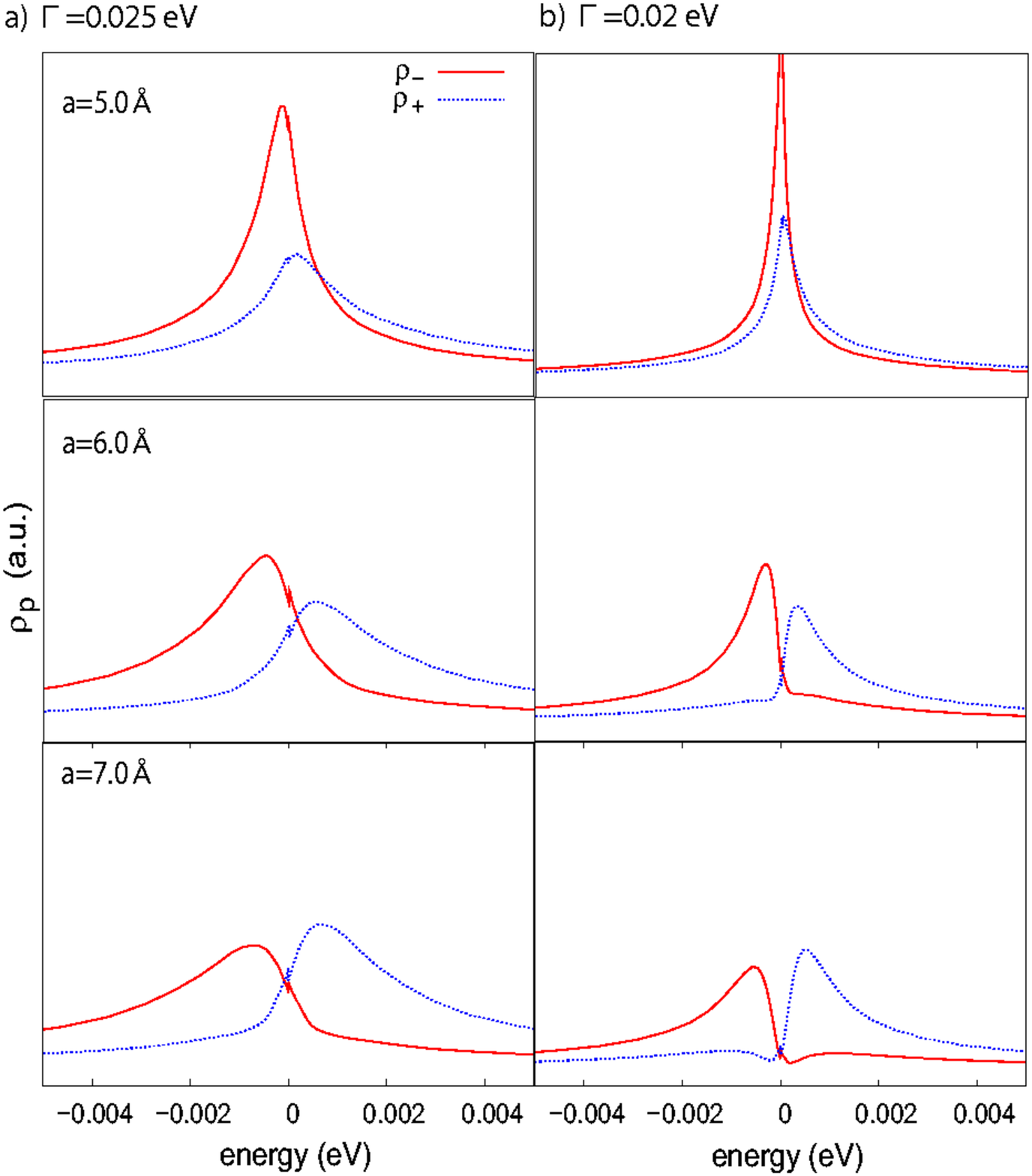}
  \caption[labelInTOC]{Single electron excitation spectra of adatom electrons
  for each parity channel.}
  \label{fig:split.eps}
\end{center}
\end{figure}
We find that the ``parity splitting'' of the single electron
excitation spectra is the origin of the dip structure in the {\it
dI/dV} spectra. As shown in \Eref{NRGform}, the operators in the model
Hamiltonian are indexed by the parity $q$. Thus, the obtained physical
properties such as single electron excitation spectra are divided with respect to the parity.
The {\it dI/dV} spectra are proportional to the average of the single electron
excitation spectra of adatom electrons on each parity channel
($\rho_+$,$\rho_-$). As shown in \Fref{fig:split.eps}, $\rho_+$ and $\rho_-$ have different peak
positions. The asymmetry of the spectra results from the difference in the
coupling between the adatom and conduction band in each parity channel (See \Eref{NRGform} and \Eref{gamma_pm}).
Previous studies shows that the parity
splitting is one of the signatures of the smearing out of the critical point due
to the breaking of the e-h symmetry \cite{Affleck1995,Sakai1992a}. In the present
study, based on the model Hamiltonian (\ref{H_gen}), the tunneling matrix
element between the localized d-electrons and the metal surface conduction electrons
has energy dependence, which breaks the e-h symmetry.
\begin{figure}[htp]
  \begin{center}
    \includegraphics[keepaspectratio=true,height=110mm]{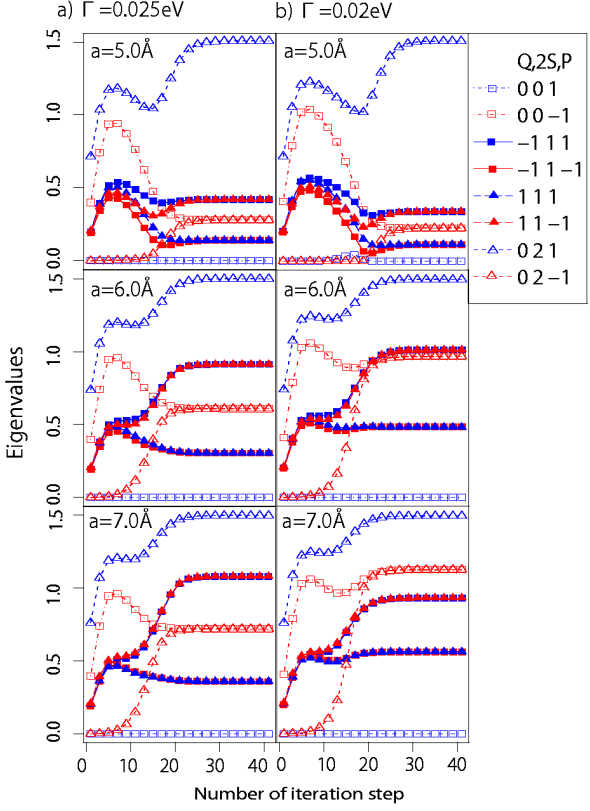}
  \end{center}
  \caption{Flows of the low-lying many-particle energy levels in odd
  iterations. $N+1$ corresponds to the iteration step number. In this
  calculation we calculate at $a=5.0, 6,0,7,0$\AA a) $\Gamma=0.025$eV
  b)$\Gamma~0.02$eV. The states are labeled by the quantum numbers total charge $Q$,
  total spin $S$, and total parity $P$.  }
  \label{fig:EFL.eps}
\end{figure}

~~In order to investigate the electron state, we plotted the
flow of low lying many-particle energies in the NRG calculation with several adatom separations
(\Fref{fig:EFL.eps}).
The states are labeled by the quantum numbers total charge $Q$,
  total spin $S$, and total parity $P$. Here we show the flows of the lowest
  state for (Q,2S,P)=(0,0,1),(1,1,1),(-1,1,1),(1,1,-1),(-1,1,-1),(0,2,1) and (0,2,-1).
 The Q=0,2S=0,P=1 state is the ground state. Q=$\pm1$,2S=1and P=$\pm1$
 states are the single electron (hole) excited states. When $a=5.0$\AA, as $N$
 increases, the energy of ($\pm 1$, 1, $\pm 1$) state decreases relatively.
 This indicates the energy gain from the formation of a singlet between
 the adatom localized spin and excited state of the conduction electron, i.e.,
 the Yosida-Kondo singlet. However, when $a$ comes close to 7.0\AA, the
 energy of the single particle excited states do not decrease so much
 but keep considerable value in large N.
 \begin{figure}[htp]
\begin{center}
  \includegraphics[keepaspectratio=true, height=50mm]{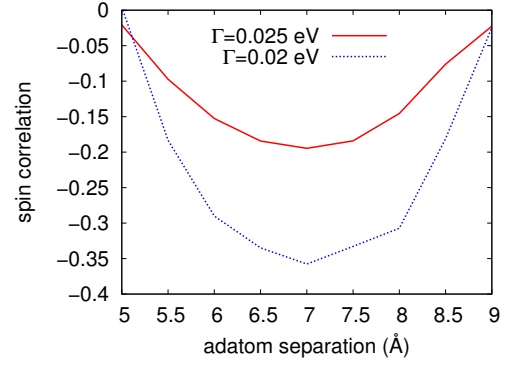}
  \caption[labelInTOC]{Spin correlation function as a function of adatom
  separation. }
  \label{spincorr}
\end{center}
\end{figure}
From the calculation result for the spin correlation function between adatom
localized spin(\Fref{spincorr}), we conclude that this large energy of single particle
excitation states originates from the interruption of the Yosida-Kondo singlet formation by the strong
antiferromagnetic correlation between the localized spins. In our
model, the e-h symmetry breaking connect the antiferromagnetic region and the
Kondo region continuously and the ground state would be the hybrid of the
localized spin-spin singlet state and the Yosida-Kondo singlet state. As the
antiferromagnetic correlation between localized spin becomes large, the
localized spin-spin singlet state would become dominant and suppress the
excitation at the Fermi level, which results in the dip structure of {\it
dI/dV}. Thus, the dip structure in {\it dI/dV} would enable us to detect the strong
antiferromagnetic correlation between localized spins.

 \begin{figure}[htp]
  \begin{center}
    \includegraphics[keepaspectratio=true,height=60mm]{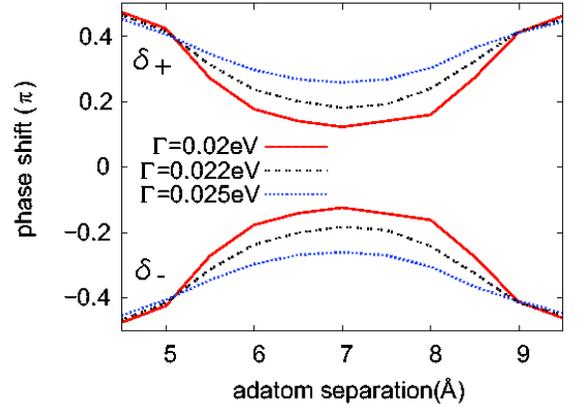}
  \end{center}
  \caption{Phase shift as a function of adatom separation. (in units of
  $\pi$ radians)}
  \label{fig:phaseshift.eps}
\end{figure}
 We can also discuss about the fixed point Hamiltonian and the phase shift
estimated from that. As shown in \Fref{fig:EFL.eps}, the energy levels converge
to some values for large N. This means that the Hamiltonian becomes close to the
fixed point Hamiltonian of renormalization transformation\cite{Krishna1980,Krishna1980a,Bulla2008}.
In this calculation, we can define the fixed point Hamiltonian as follows;
\begin{eqnarray}
H_{eff}^{N}&=&\Lambda^{(N-1)/2}\bigl\{\sum^{N-1}_{q=\pm,n=0}\eta _{qn}(f^\dagger_{qn} f_{qn}+ H.C.) \nonumber \\
&+&K_{q}f^\dagger_{0}f_{0} \bigr\} \label{effectiveH}
\end{eqnarray}
 \noindent $q=\pm$ is the parity index. The second term with $K_q$ is
 related to the influence of potential scattering in each channel. The strength of the potential scattering $K_q$ and the phase
 shift $\delta_q$ have the following relation:
\begin{eqnarray}
\delta_q=-\tan^{-1}(\pi\rho K_q)
\end{eqnarray}
We estimate $K_q$ in each conduction channel by comparing the many-particle
energy of first excited state in the last iteration of NRG calculation with that
of Eq.(\ref{effectiveH}). \\
~~The phase shift is related to the number of
electrons and holes which are virtually bounded by localized spin -- i.e., the
phase shift is a barometer of the existence of the Kondo effect. If $\delta_\pm=\pi/2$, it means that one
electron-hole pair is bounded to each localized spin and quenches it individually (i.e,. two Yosida-Kondo singlets
are formed). On the other hand, $\delta_\pm=0$ indicates that a localized
spin-spin singlet is formed \cite{Sakai1992a}. In \Fref{fig:phaseshift.eps}, we
show the result of the phase shift calculation as a function of adatom
separation with several values of $\Gamma$.\\
~~The phase shift changes gradually
and has an intermediate value between $\pm\pi/2$ to 0 (Fig.
\ref{fig:phaseshift.eps}). The smallest value of $\delta_\pm$ becomes close to
0 as $\Gamma$ decreases. This result also indicates that the dip structure
around $a=7.0$\AA~ originates from the dominance of the spin-spin singlet state in the ground
state. Though the values of $\delta_\pm$ at each adatom separation are different
by $\Gamma$, the change of $\delta_\pm$ is smooth in all cases. If a
critical point separates the Kondo region and the antiferromagnetic region,
the possible values of $\delta_\pm$ are only 0 or $\pm\pi/2$
\cite{Sakai1992a,Sakai1992,Silva1996} .
The intermediate value of $\delta_\pm$ would result from the crossover
transition and supports the conclusion that there is no critical point in the
system of a magnetic dimer on a metal surface.

\section{Summary}
To investigate how the transition between the Kondo effect dominant region and the
antiferromagnetic RKKY interaction dominant region can be observed through
scanning tunneling spectroscopy (STS), we
calculate the differential conductance ({\it dI/dV}) corresponding to STS
measurements for two magnetic atoms adsorbed on a metal surface with the aid of
the numerical renormalization group technique. We find that the peak structure of the {\it
dI/dV} spectra changes gradually as a function of the adatom separation and the
coupling ($\Gamma$) between the adatoms and the metal surface conduction band.
When $\Gamma$ becomes small, the peak disappears and a dip structure appears near the Fermi level. This dip
structure originates from the parity splitting of the single electron excitation spectra and the manifestation of the
strong antiferromagnetic correlation between the localized spins. The result of
the phase shift calculation supports the conclusion that there is no critical
point between the Kondo effect dominant region and the antiferromagnetic RKKY
interaction dominant region but a crossover transition connect these regions.
\\ ~~In conclusion, we show that the crossover transition from the Kondo
region to the antiferromagnetic region in two-impurity Kondo effect can be observed through the change of the STS spectra.
In particular, the existence of the strong antiferromagnetic correlation between
localized spins are observed as dip structures in the {\it dI/dV}.
Our results
indicate the possibility in STM observation with the resolution of magnetic
interactions on surface system, which would contribute to the realization of spintronics.\\

\section{Acknowledgments}
  This work is supported by the Ministry of Education, Culture, Sports, Science and Technology of
  Japan (MEXT) through their Special Coordination Funds for the Global Center of Excellence (GCOE)
  program (H08) ``Center of Excellence for Atomically Controlled Fabrication
  Technology", Grant-in-Aid for Scientific Research (C)(19510108); and the New Energy and Industrial Technology Development
  Organization (NEDO).  E. Minamitani would like to thank the NIHON L'OREAL K.K. and Japan Society
  for the Promotion of Science (JSPS) for financial support and H. Matsuura (Osaka Univ.) for helpful
  discussions. Some of the calculations presented here were performed using the computer facilities
  of Cyber Media Center (Osaka University), the Institute of Solid State Physics (ISSP Super Computer
  Center, University of Tokyo), and the Yukawa Institute (Kyoto University).

\end{document}